# Design method for quasi-isotropic transformation materials based on inverse Laplace's equation with sliding boundaries

**Zheng Chang[1], Xiaoming Zhou[1], Jin Hu[2] and Gengkai Hu [1*]**

[1]*School of Aerospace Engineering, Beijing Institute of Technology, Beijing 100081, P. R. China*
[2]*School of Information and Electronics, Beijing Institute of Technology, Beijing 100081, P. R. China*
[*]*Corresponding author: hugeng@bit.edu.cn*

**Abstract:** Recently, there are emerging demands for isotropic material parameters, arising from the broadband requirement of the functional devices. Since inverse Laplace's equation with sliding boundary condition will determine a quasi-conformal mapping, and a quasi-conformal mapping will minimize the transformation material anisotropy, so in this work, the inverse Laplace's equation with sliding boundary condition is proposed for quasi-isotropic transformation material design. Examples of quasi-isotropic arbitrary carpet cloak and waveguide with arbitrary cross sections are provided to validate the proposed method. The proposed method is very simple compared with other quasi-conformal methods based on grid generation tools.

## 1. Introduction

The form invariance of Maxwell's equations under coordinate transformations constructs the equivalence between the geometrical space and material space, the developed transformation optics theory [1,2] is opening a new field that may cover many aspects of physics and wave phenomenon. Transformation optics (TO) makes it possible to investigate celestial motion and light/matter behavior in laboratory environment by using equivalent materials [3]. Illusion optics [4] is also proposed based on transformation optics, which can design a device looking differently. The attractive application of transformation optics is the invisibility cloak, which covers objects without being detected [5]. Most of recent researches have been conducted to resolve the following three challenges concerning design of cloaks. The first challenge is how to design a cloak with complex shapes. The works on this direction include the studies on complex but regular shapes [6-11], angular Fourier expansions for an arbitrary shape [12] and arbitrary cloaks designed by solving Laplace's equation [13]. The second challenge is the singularity of the transformed material parameters at the inner boundaries for two-dimensional (2D) cloaks. The material singularity poses a significant problem for practical realization of the designed cloak. One can overcome this second problem by projecting on a mirror-symmetric cross section from a three-dimensional (3D) cloak [14], by adjusting the out-of-plane stretches for arbitrary 2D cloaks [15], or eventually by using non-Euclidean geometry theory [16]. The third challenge is about the frequency band of operation. The transformation materials are usually anisotropic and they involve material parameters not easily found in nature. Good candidates are metamaterials, a kind of composite materials whose effective properties are very anomalous due to the microstructure resonance. The resonance inevitably limits the frequency band of operation and is often accompanied by large dissipations. So efforts are now made for transformation materials realized by simple metamaterials or even by conventional materials, thus confronting the third challenge: how to design broadband transformation materials with isotropic and easily realizable material parameters. Li and Pendry [17] make the first step on this issue by designing quasi-isotropic carpet cloaks, they use grid generation technique to find appropriate quasi-conformal transformations for the designed devices. Based on the method proposed by Li and Pendry [17], Kallos *et al.* [18] proposed a simplified method to design quasicloaks which are more easily implemented in practice. Successful fabrications of broadband carpet cloaks in microwave and optical frequencies accelerate the development of this field [19-21].

Recently, Hu *et al.* [13] found the equivalence between coordinate transformation and spatial deformation, they propose to use Laplace's equation to determine the deformation of coordinate grids during the transformation, and in turn to design arbitrary cloaks. In addition, the singularity of arbitrary 2D cloaks [15] can be easily removed in context of this method. Since the mapping based on Laplace's equation together with Dirichlet-Neumann (sliding) boundary is quasi-conformal[22], and it is also shown by Li and Pendry that quasi-conformal maps minimize material anisotropy, so it is interesting to examine Laplace's equation with sliding boundary for quasi-isotropic material design in case of transformation optics. This is the objective of the present work. The paper is arranged as follows, in Sec. 2, we firstly recall some basic results of transformation optics, and then we will explain that the inverse Laplace's equation corresponds to extremum condition of Winslow functional, which minimizes also the material anisotropy, finally we point out that the sliding boundary is a necessary condition for quasi-conformal transformation. Examples and detailed discussions are given in Sec. 3, and followed by some conclusions in Sec. 4.

## 2. Quasi-conformal transformation based on inverse Laplace's equation

*2.1 Preliminary*

Consider a mapping $\mathbf{x}'=\mathbf{x}'(\mathbf{x})$, which maps every point $\mathbf{x}$ in $\Omega$ uniquely to $\mathbf{x}'$ in a new space $\Omega'$, according to the transformation optics [1], the transformation material in the transformed space $\Omega'$ is related to its original flat space $\Omega$ by

$$\boldsymbol{\varepsilon}' = \frac{\mathbf{A}\varepsilon_0\mathbf{A}^T}{\det\mathbf{A}}, \tag{1a}$$

$$\boldsymbol{\mu}' = \frac{\mathbf{A}\mu_0\mathbf{A}^T}{\det\mathbf{A}}, \tag{1b}$$

Where $\mathbf{A}$ ($A_{ij}=\partial x_i'/\partial x_j$) is the Jacobian tensor. It is useful to interpret $\mathbf{x}'$ as another Cartesian coordinate superposed on $\mathbf{x}$, then the mapping defines a deformation field on the original space $\Omega$, characterized by the deformation gradient tensor $\mathbf{A}$. The tensor $\mathbf{A}$ can be decomposed into a pure stretch deformation (described by a positive definite symmetric tensor $\mathbf{V}$) and a rigid-body rotation (described by a proper orthogonal tensor $\mathbf{R}$), $\mathbf{A}=\mathbf{VR}$, we get $\mathbf{AA}^T=\mathbf{VV}^T$ in Eq. (1). In the local principal system of the tensor $\mathbf{V}$, if the pure stretches are denoted by $\lambda_i$, which are the eigenvalues of $\mathbf{V}$, Eq. (1) can be rewritten as [13]

$$\boldsymbol{\varepsilon}' = \varepsilon_0 \mathrm{diag}[\frac{\lambda_1}{\lambda_2\lambda_3}, \frac{\lambda_2}{\lambda_1\lambda_3}, \frac{\lambda_3}{\lambda_1\lambda_2}], \tag{2a}$$

$$\boldsymbol{\mu}' = \mu_0 \mathrm{diag}[\frac{\lambda_1}{\lambda_2\lambda_3}, \frac{\lambda_2}{\lambda_1\lambda_3}, \frac{\lambda_3}{\lambda_1\lambda_2}]. \tag{2b}$$

Equation (2) implies that the transformation materials are closely related to the deformation fields during the mapping. It is also found that the dilatational and shear deformations will determine respectively the magnitude and anisotropy of the transformation material.

*2.2 Variational form of inverse Laplace's equation*

Laplace's equation [13] has been introduced to calculate the transformation (or deformation) during the mapping

$$\nabla_{\mathbf{x}}^2 \mathbf{x}' = 0. \tag{3}$$

In Eq. (1), the material parameters are defined in the transformed space. However to facilitate numerical computation, it is better to use the inverse Laplace's equation to give the solution of the deformation field in the transformed space

$$\nabla_{\mathbf{x}'}^2 \mathbf{x} = 0. \tag{4}$$

Geometrically, equation (4) means the inverse spatial transformation, i.e., the transformation from the transformed space to its original space. For an arbitrary-shaped transformation material, its original space usually has a regular shape, thus the boundary conditions corresponding to Eq. (4) can be easily imposed.

Now we will examine the characteristic of the transformation determined by the inverse Laplace's equation. Without loss of generality, consider a 2D problem, equation (3) is in fact the Euler-Lagrange equation of the following length functional [23]

$$F_L(\mathbf{x}') = \frac{1}{2}\int_{\Omega}\left(g_{11} + g_{22}\right) d\mathbf{x}, \tag{5}$$

where the covariant metric is defined by $g_{ij} = [\mathbf{V}\mathbf{V}^T]_{ij}$ . In other words, Laplace's equation (3) results in minimized values of the length functional (5). The length functional for the inverse Laplace's equation (4) is given by

$$F_W(\mathbf{x}) = \frac{1}{2}\int_{\Omega'}\left(g^{11} + g^{22}\right) d\mathbf{x}'. \tag{6}$$

where $g^{11} = g_{11}/g$ , $g^{22} = g_{22}/g$ are components of the contravariant metric, $d\mathbf{x} = d\mathbf{x}'/\sqrt{g}$ , and $g$ is the determinant of the metric tensor $g_{ij}$ . Converting Eq. (6) from the original space to the transformed space, we have the following result

$$F_W(\mathbf{x}') = \frac{1}{2}\int_{\Omega}\frac{g_{11} + g_{22}}{\sqrt{g}} d\mathbf{x}, \tag{7}$$

Equation (7) is known as the Winslow functional [23], corresponding to smoothness for the deformation field induced by the mapping. Physical insights can be further discovered in the local principle system, where $g_{11} = \lambda_1^2$ , $g_{22} = \lambda_2^2$ and $\sqrt{g} = \lambda_1 \lambda_2$ . With the principle stretch $\lambda_i$ , the Winslow functional becomes

$$F_W(\mathbf{x}') = \frac{1}{2}\int_{\Omega}\left(\frac{\lambda_1}{\lambda_2} + \frac{\lambda_2}{\lambda_1}\right) d\mathbf{x}. \tag{8}$$

So the inverse Laplace's equation determines the transformation that minimize this functional and will lead to the optimal result $\lambda_1 \approx \lambda_2$ for all infinitesimal elements. It is seen from Eq. (2) that transformation materials are quasi-isotropic under this mapping, if $\lambda_3 = 1$ .

Li and Pendry [17] obtain quasi-conformal mappings by a grid generator based on the modified Liao's functional

$$F_{ML}(\mathbf{x}') = \frac{1}{2}\int_{\Omega}\frac{\left(g_{11} + g_{22}\right)^2}{g} d\mathbf{x}. \tag{9}$$

In the local principle system, above functional is rewritten as

$$F_{ML}(\mathbf{x}') = \frac{1}{2}\int_{\Omega}\left(\frac{\lambda_1}{\lambda_2} + \frac{\lambda_2}{\lambda_1}\right)^2 d\mathbf{x}. \tag{10}$$

Comparing Eq. (8) and Eq. (10), we find that the integrand of $F_{ML}$ is just the square of the Winslow integrand. From the point of grid generations, there is almost no difference between these two functionals, except that the "Modified Liao" grids are able to keep from folding [22]. It is worth to note that the folding of coordinate system often takes place in very complex cases, seldom observed in transformation optics. As shown by equation (8), the inverse Laplace equation (4) has a property to minimize the material anisotropy. However, in order to get a quasi-conformal mapping, sliding boundary condition is also necessary, as pointed by Thompson *et al* [22].

*2.3 Sliding boundary conditions*

The transformation material parameters near the boundaries are mainly dominated by imposed boundary conditions. In transformation optics, the boundary conditions prescribe the functionality of devices. For complete cloaks [1], the outer boundary is fixed to ensure the grid continuity and omnidirectional invisibility. However for carpet cloaks and some other directional devices, we can release the fixed boundary without changing the functionality by imposing the Neumann boundary conditions $\partial x_i / \partial \mathbf{n}' = 0$, where $x_i$ is the variable whose isoline is orthogonal to the considered boundary, and $\mathbf{n}'$ is the unit vector normal to the boundary of interest. Note that the condition is written in the transformed space, corresponding to the inverse Laplace's equation (4). The Neumann boundary is also called sliding boundary, where the grids on the boundary can be adjusted along the boundary to make the adjacent grids tend to be square. The inverse Laplace equation together with sliding boundary condition yields quasi-conformal mapping, it can be used as an alternative to attain quasi-isotropic transformation media without aid of grid generation theory as in reference [17].

In summary, we suggest to use directly the inverse Laplace's equation $\nabla^2_{\mathbf{x}'}\mathbf{x} = 0$ together with the sliding boundary condition $\partial x_i / \partial \mathbf{n}' = 0$ to determine quasi-conformal transformation, and to design quasi-isotropic transformation media. In the following section, examples of designing quasi-isotropic arbitrary carpet cloak and waveguide are given to illustrate the proposed method.

## 3 Applications and discussions

*3.1 Carpet cloaks*

We firstly examine the carpet cloak proposed by Li and Pendry [17] by using the proposed method. The virtual system is compressed at the sliding edge attached to the ground plane to leave enough space for an object to be concealed. The exact conformal transformation for carpet cloaks cannot be realized, since the total spatial compression ratio along the horizontal and vertical directions in the physical systems are different. That means the in-plane permeabilities will be slightly anisotropic for transverse electric (TE) detecting waves. A simplification will be made here by replacing the permeability with unity to get isotropic and pure dielectric carpet cloaks. In contrast, Li and Pendry [17] enable all edges to be slipping and then truncate the in-plane permeabilities. Numerical simulations reveal that the latter has better cloaking effects. For comparison, we calculate the carpet cloak proposed by Li and Pendry [12] by the inverse Laplace's equation with the sliding boundary conditions imposed on all edges. The COMSOL multiphysics is used for numerical computation, which enables the integration for calculation of material parameter and validation in a two-step model. The anisotropic factor of the present cloak ranges from 1.04 to 1.043, very close to 1.04 optimized by the modified Liao's functional [17]. The coordinate grids in the transformed space are shown in Fig. 1(a) with color maps for profile of out-of-plane permittivity. Figure 1(b) shows the electric field pattern for a Gaussian beam of frequency 0.8GHz incident at $45^\circ$ on the carpet cloak. Compared with the bare object (PEC boundary) without cloak shown in Fig.1(c) and a conducting ground plane in Fig.1 (d), it is seen that the designed carpet cloak indeed reflects the detecting beam like a conducting ground plane.

The proposed method is very simple to use, in order to design quasi-isotropic transformation materials, it suffice to solve the inverse Laplace's equation with sliding boundary condition without use of grid generation theory. Figure 2 gives another example of designing a quasi-isotropic arbitrary carpet cloak, where the arbitrary boundary (PEC) can be used to fit the irregular shape of concealed objects. Notice that the sliding boundary conditions are imposed on all edges. Calculation results show that the anisotropic factor is about 1.055. After the truncation, the out-of-plane permittivity of the quasi-isotropic and dielectric cloak ranges from 0.2 to 4, as shown in Fig. 2(a). Fig. 2(b) shows the electric field distribution when the same Gaussian beam used in Fig. 1 impinges the dielectric cloak. Compared with the object without the cloak shown in Fig. 2(c) and a conducting ground plane in Fig. 2(d), we find that the ground plane reflections can be observed outside the cloaking material, this clearly demonstrates the excellent performance of the designed carpet cloak.

In designing the previous carpet cloaks, the back ground media is assumed to be free space, so metamaterial has to be employed. The metamaterials with local resonant microstructure proposed by Liu *et al.*[19] can be used for such purpose. However, this inevitably limits the frequency band of operation. To overcome this difficulty, the method proposed by Kallos *et al.* [18] can be used to simplify the designed carpet cloaks with conventional materials.

*3.2 Arbitrary waveguide*

The waveguide presented here is a directional device that can guide waves along a desired path. Figure 3 shows the scheme of the boundary settings for a waveguide with arbitrary cross sections. An example is given in Fig. 4 to demonstrate that the arbitrary waveguide is easily designed by using the proposed method. To get quasi-conformal mappings, the virtual space is a $2\mathrm{m}\times 27.27\mathrm{m}$ rectangle so that the virtual and material spaces have the same areas. In this case, the anisotropic factor ranges from 1 to 1.02. Figure 4(a) gives the pattern of out-of-plane permittivities after the truncation of in-plane permeabilities as well as the transformation grids in black. When a Gaussian beam of an operation frequency 0.5GHz enters the waveguide, it will experience the expanding, squeezing, and bending before transmitted to a different place, as shown in Fig. 4(b) for the electric field distribution. There are no reflections and signal losses during the transmission. Waves travel in the designed waveguide, just like in a hollow waveguide.

**4. Conclusions**

Based on the fact that the inverse Laplace's equation together with sliding boundary gives a quasi-conformal mapping and that a quasi-conformal mapping minimizes the transformation material anisotropy, we propose to use directly inverse Laplace's equation with sliding boundary for quasi-isotropic transformation material design. For demonstration, quasi-isotropic arbitrary carpet cloak and waveguide are designed, both showing the excellent performance. The broadband applications can be anticipated due to their isotropic nature. Compared with other quasi-conformal methods based on grid generation techniques [17, 24], the proposed method is very simple, the quasi-isotropic transformation materials can be determined by solving directly inverse Laplace's equation with sliding boundary without grid generation theory. The same idea can also be applied to the design of acoustic cloaks according to the method developed in the reference [25].

**Acknowledgments**

The authors would like to thank the anonymous reviewers for their valuable suggestions. This work is supported by the National Natural Science Foundation of China (10702006, 10832002), and the National Basic Research Program of China (2006CB601204).

## References and links:

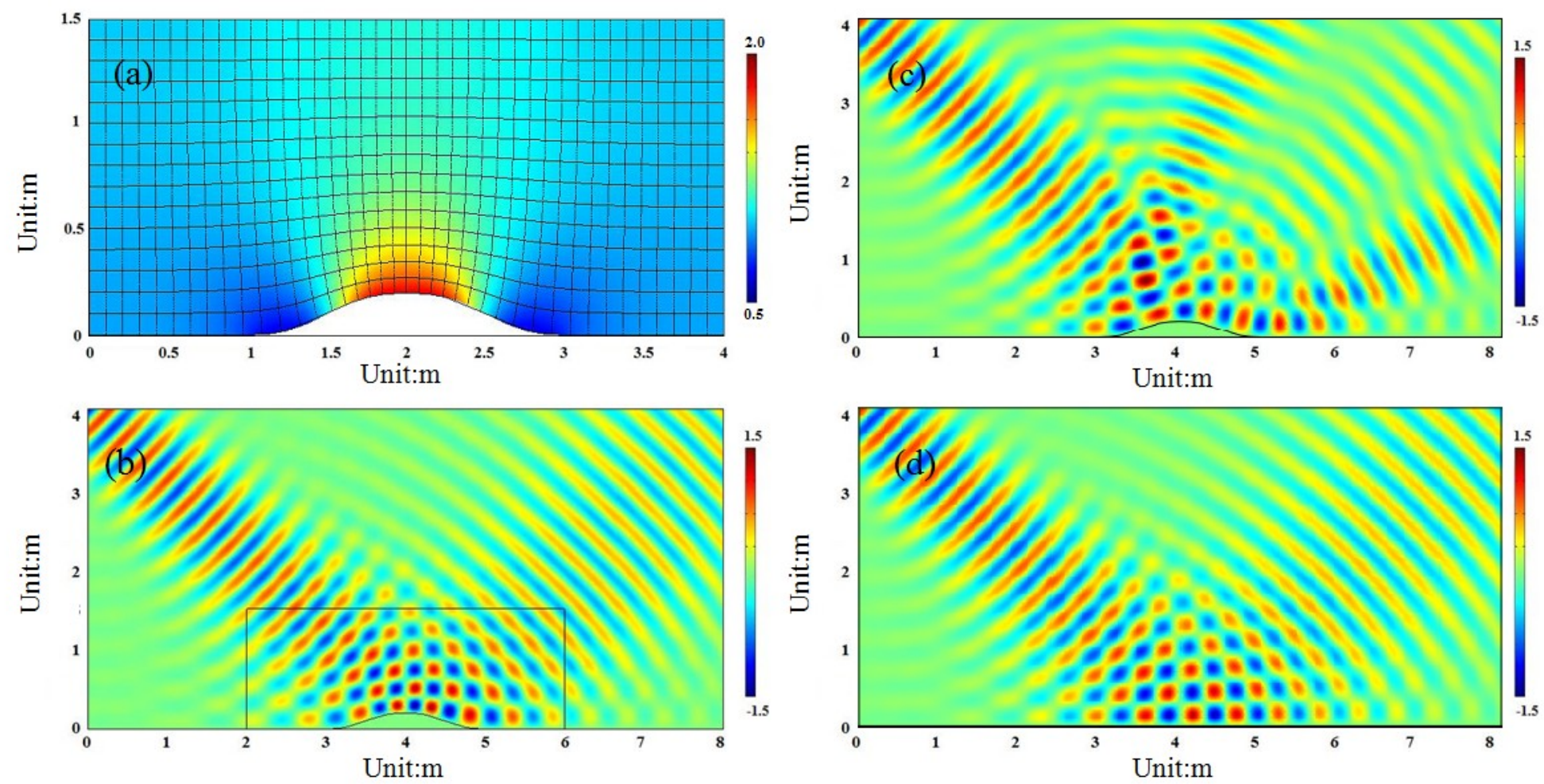

Fig.1. (a) The out-of-plane permittivity of dielectric carpet cloak for TE waves with transformation grids in black, (b) the electric field pattern for a Gaussian beam of frequency 0.8GHz impinging the cloak, (c) the electric field pattern of a PEC object without cloak and (d) the electric field pattern of a conducting ground plane.

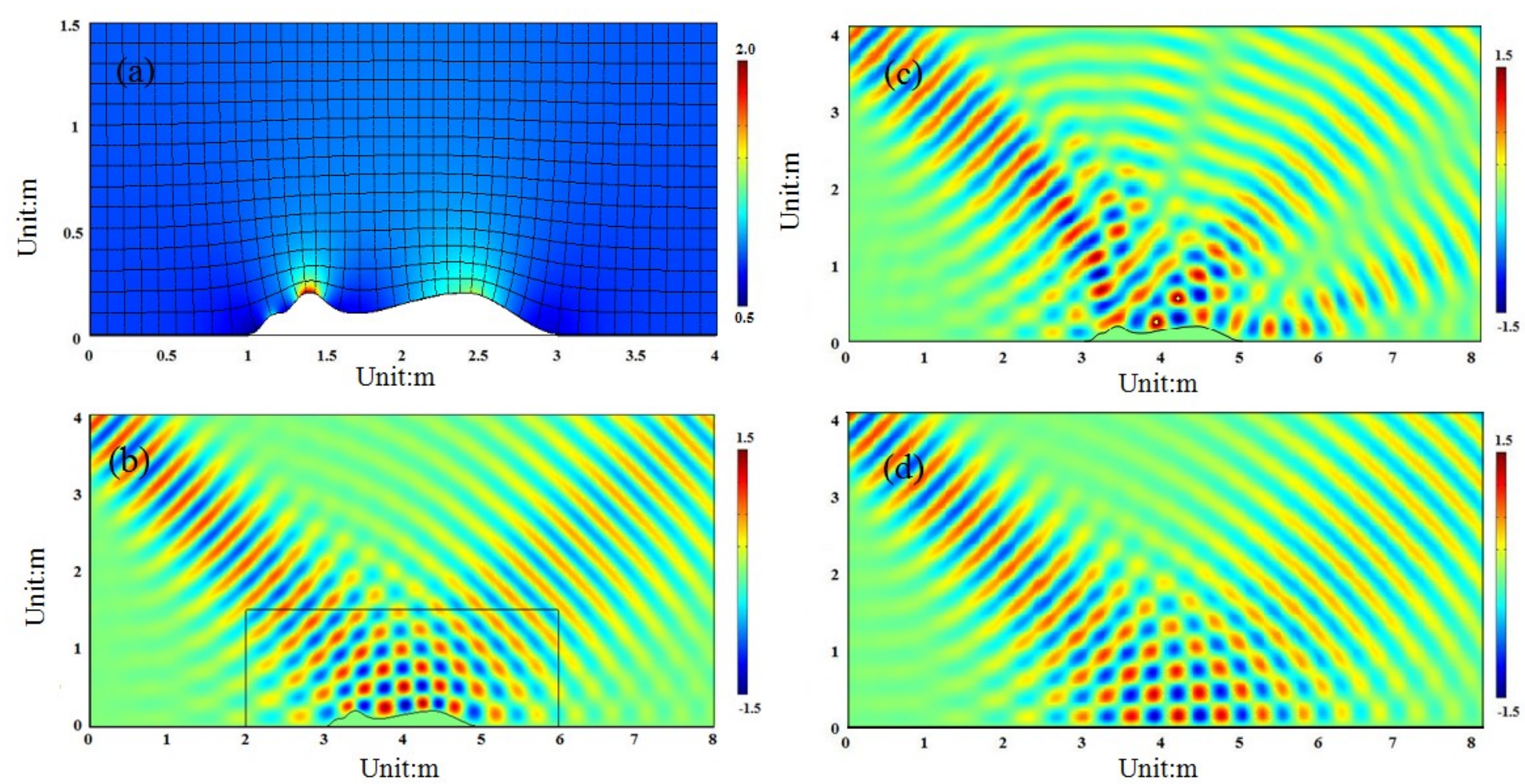

Fig.2. (a) The out-of-plane permittivity of an arbitrary carpet cloak for TE waves with transformation grids in black, (b) the electric field pattern for a Gaussian beam of frequency 0.8GHz impinging the cloak, (c) the electric field pattern of a PEC object without cloak and (d) the electric field pattern of a conducting ground plane.

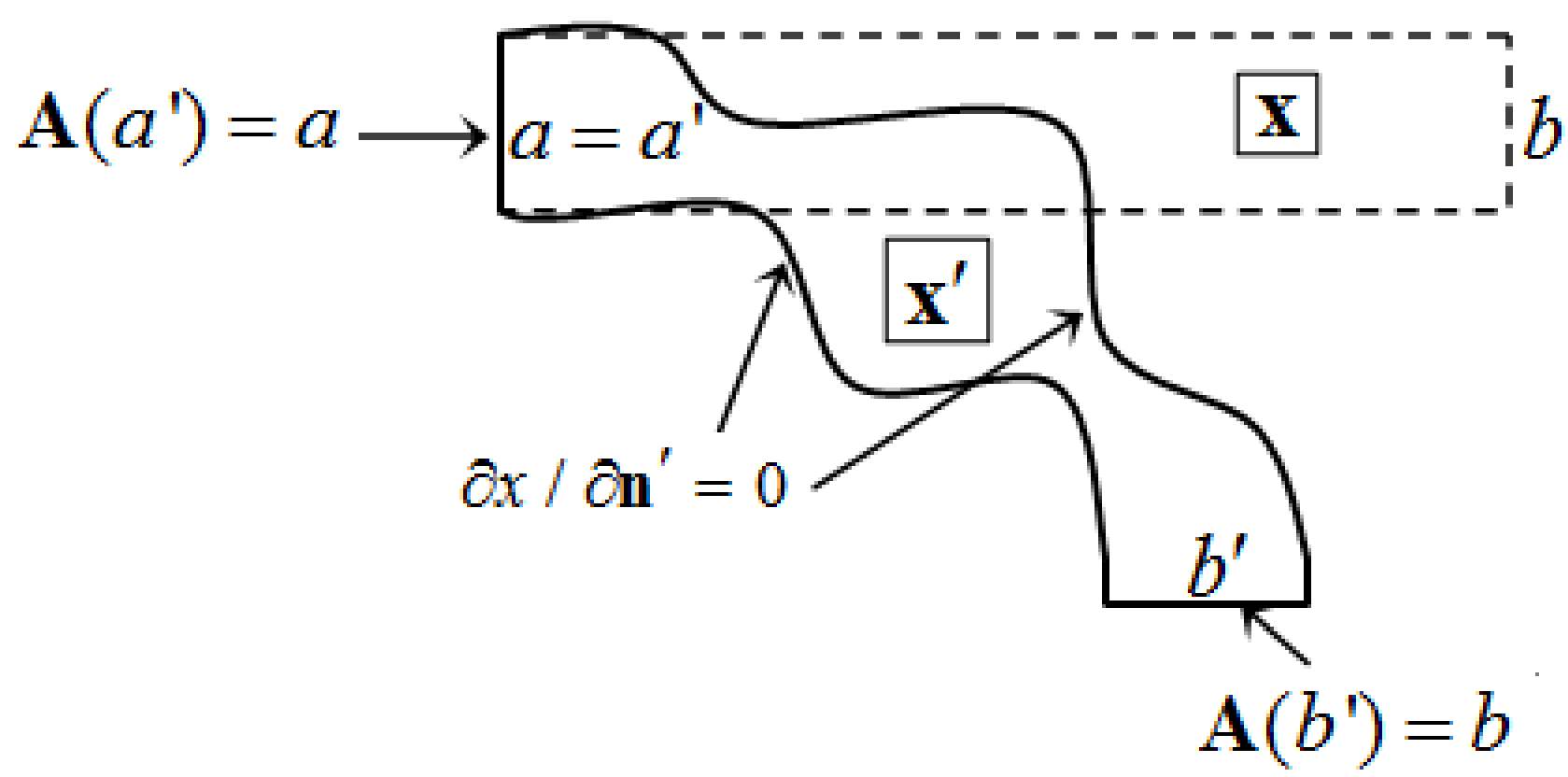

Fig.3. The scheme of the boundary settings for an arbitrary waveguide. $\mathbf{A}(a') = a$ and $\mathbf{A}(b') = b$ indicate the correspondences between the boundaries $a$ and $b$ before and after the transformation.

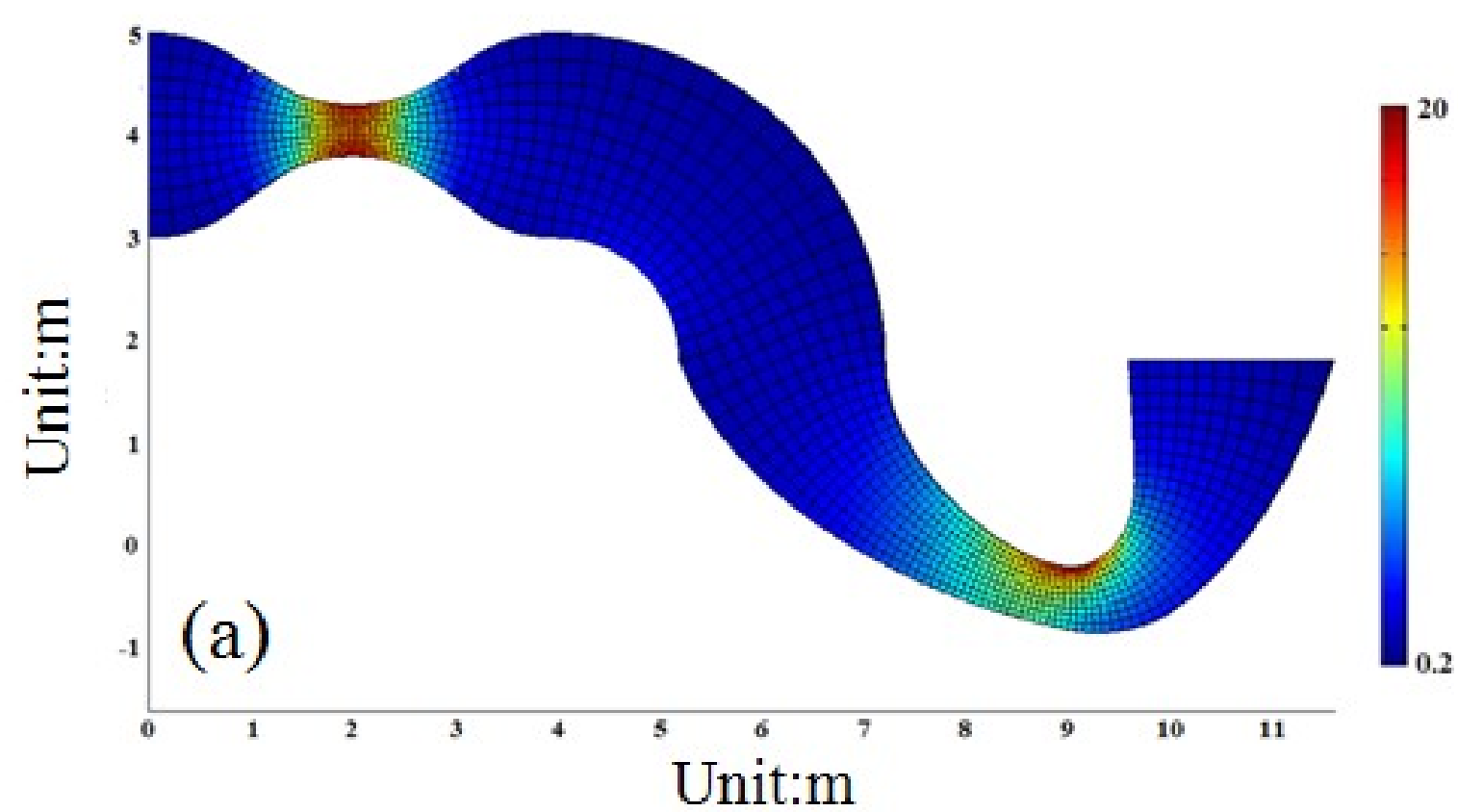

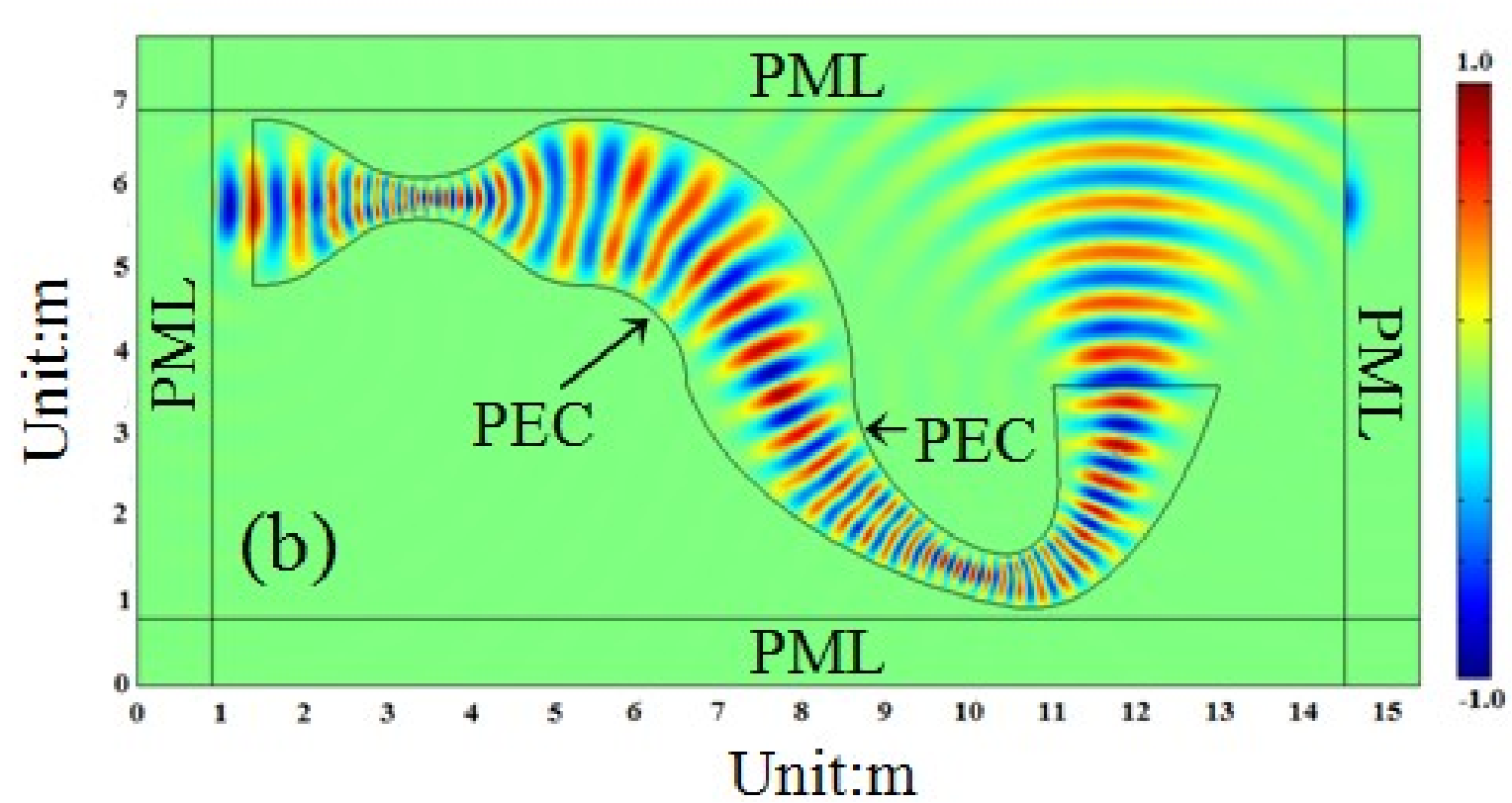

Fig.4. (a) The out-of-plane permittivity of a dielectric waveguide of arbitrary cross sections with transformation grids in black, and (b) the electric field pattern for a Gaussian beam of frequency 0.5GHz propagating in the waveguide.